\documentclass[aps,prd, reprint, showpacs,superscriptaddress,floatfix,amsmath,amssymb]{revtex4-1}

\usepackage{lipsum}
\usepackage{multirow}
\usepackage{graphicx}
\usepackage{bm}
\usepackage{color}

\begin{document}

\date{\today}

\title{Spacecraft clocks and relativity: Prospects for future satellite missions}

\author{Raymond Ang\'elil}\email{rangelil@physik.uzh.ch}
\affiliation{Institute for Computational Science, Universit\"at
  Zurich,Winterthurerstrasse 190, CH-8057 Z\"urich, Switzerland}

\author{Prasenjit Saha}
\affiliation{Department of Physics, Universit\"at Z\"urich,Winterthurerstrasse 190, CH-8057 Z\"urich, Switzerland}

\author{Ruxandra Bondarescu}
\affiliation{Department of Physics, Universit\"at Z\"urich,Winterthurerstrasse 190, CH-8057 Z\"urich, Switzerland}

\author{Philippe Jetzer}
\affiliation{Department of Physics, Universit\"at Z\"urich,Winterthurerstrasse 190, CH-8057 Z\"urich, Switzerland}

\author{Andreas Sch\"arer}
\affiliation{Department of Physics, Universit\"at Z\"urich,Winterthurerstrasse 190, CH-8057 Z\"urich, Switzerland}

\author{Andrew Lundgren}
\affiliation{Albert Einstein Institut f\"ur Gravitationsphysik, Callinstrasse 38, 30167 Hannover, Germany}

\begin{abstract}
The successful miniaturization of extremely accurate atomic clocks invites prospects for satellite missions to perform precise timing experiments.
This will allow effects predicted by general relativity to be detected in Earth's gravitational field. 
In this paper we introduce a convenient formalism for studying these effects, and compute the fractional timing differences generated by them for the orbit of a satellite capable of accurate time transfer to a terrestrial receiving station on Earth, as proposed by planned missions. We find that (1) Schwarzschild perturbations will be measurable through their effects both on the orbit and on the signal propagation, (2) frame-dragging of the orbit
will be readily measurable, and (3) in optimistic scenarios, the spin-squared metric
effects may be measurable for the first time ever.  Our estimates suggest that a clock with a fractional timing inaccuracy of 
$10^{-16}$ on a highly eccentric Earth orbit will measure all these effects, while for a low Earth circular orbit like that of the Atomic Clock Ensemble in Space mission, detection will be more challenging.
\end{abstract}

\keywords{atomic clocks, general relativity, kerr, schwarzschild, geophysics, timing}

\maketitle

\section{Introduction}

Atomic clock technology has improved dramatically over the past decade. Present-day Earthbound atomic clocks boast fractional timing inaccuracies of $\sim10^{-18}$ \cite{Chou, Middelmann, Hinkley}. The most stable space-qualified clock, with stability $\sim10^{-16}$ has been built for the Atomic Clock Ensemble in Space (ACES) mission, and will be placed on the International Space Station in 2016. As we show in this paper, a variety of relativistic effects are extremely important at these accuracy levels. Optical time and frequency transfer over free space has reached a residual stability of $\sim10^{-18}$ over distances of a few kilometers \cite{Giorgetta}. As such technology improves, it is expected that highly accurate clocks on Earth will be commonly used for orbit determination. 

The theory of general relativity explains gravitation as a geometrical phenomenon arising from a curved four-dimensional spacetime. The Earth, carrying mass and momentum, determines the trajectories - both spatially and temporally - of satellites which fall freely around it. The Einstein equivalence principle tells us that regardless of the position or velocity of a freely falling satellite, its onboard clock will tick routinely in equal intervals of time in its frame. This paper simulates an experiment to test gravitational physics by capturing this notion: a satellite carries an ultraprecise atomic clock and broadcasts tick signals to an Earth-based receiving station, whose arrival times are compared with a local clock. 

\begin{figure}[!h]
\begin{center}
\includegraphics[scale =0.6,trim=2cm 10cm 10cm 1.2cm,clip=true]{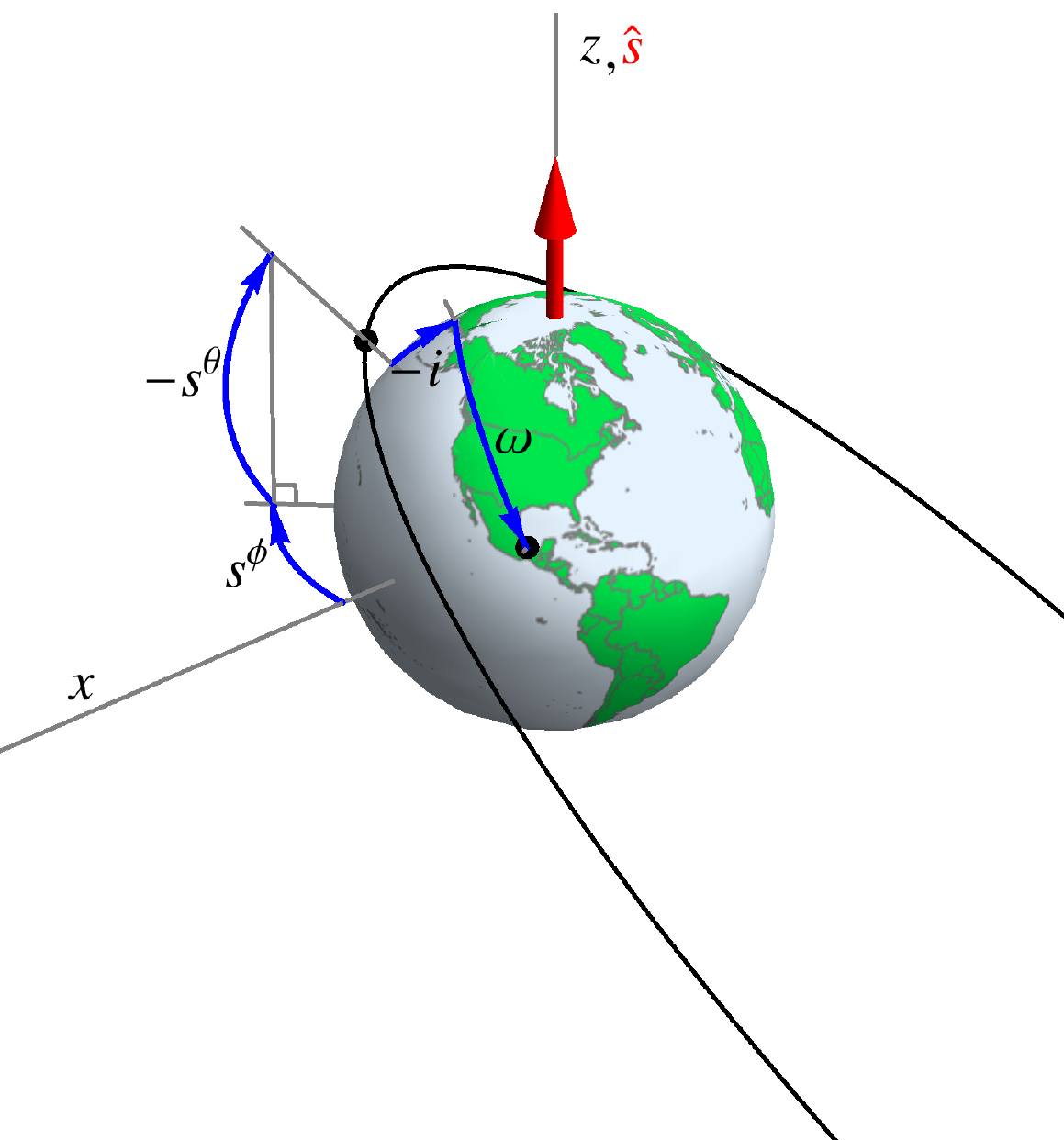}
\caption{\label{fig:orientation} Schematic of the system for an eccentric orbit. Here we define the angles in Earth's spin reference frame. The two black points mark the satellite orbit's perigee and the Earth receiving station. Two angles appear negative, as they are defined positively when measured from apogee. For computational simplicity, and without loss of generality, the Earth spin axis is always along the $z$ direction.}
\end{center}
\end{figure}

A direct consequence of the geometrical description of gravity is that a clock in a shallow gravitational field runs more rapidly than one wallowing more deeply in the field. For Earth satellites, this affects the redshift at the $G M_\oplus/(c^2 r_\oplus) \sim 10^{-9}$ level (see Fig. \ref{fig:all_signals}). Thus a fractional timing stability of $10^{-16}$ implies a sensitivity to gravitational time dilation of $10^{-7}$. We show that at these remarkable accuracy levels, higher-order relativistic effects must be taken into account.

An experiment with a freely falling satellite, which periodically sends tick signals to an Earth receiving station, is sensitive to the full four-dimensional trajectory. This is in contrast to Earth-based fixed clock experiments, which probe gravity purely through its effect on time, and not on space. State-of-the-art space-based atomic clocks, and modern time transfer technology promise orbit tracking to unprecedented accuracy. As the accuracy of clocks increases, a host of relativistic effects become important.

In this paper we address the forward problem: calculating the relativistic observables from Earth-orbiting clocks. Our calculations suggest that the Shapiro delay, Mercury-like orbit precession, frame-dragging, and possibly even spin-squared effects, will be detectable by future satellite timing missions. Our approach is general and applicable to a multitude of orbits around any gravitating body. An interactive computer program which calculates all the effects discussed in this paper is available as an online supplement, downloadable from the journal website. 

As examples, we will consider an eccentric orbit, as well as a low Earth circular orbit. An eccentric orbit is 
particularly advantageous because it brings the clock through various gravitational field strengths and thus permits the relativistic effects to modulate. Also, we find that the higher perigee velocities of the elliptical orbit boost the relativistic orbit effects by about an order of magnitude, but do not affect relativistic signal transmission effects. 

\section{Calculation of relativistic effects}

Two tick signals, emitted from the satellite's non-Keplerian trajectory, are separated by an interval of local proper time $\Delta t_e$. They travel on neighboring, typically bent paths to the receiving station, which measures the gap between their arrival times $\Delta t_a$ in the local proper time at the receiver. The redshift is then calculated from 
\begin{equation}\label{crux}
z = \frac{\Delta t_e}{\Delta t_a} - 1.
\end{equation}

This transfer can occur as frequently along the orbit as necessary. The clock enables this action by controlling $\Delta t_e$ of sequential emissions down to the clock accuracy. Equation (\ref{crux}) naturally includes all special relativistic effects.

To calculate the relevant trajectories, we use the standard Hamiltonian formulation for freely-falling bodies \cite{necronomicon} with the Hamiltonian given by
\begin{equation}\label{hamil}
H = {\textstyle \frac12}\,g^{\alpha\beta}\,p_{\alpha}\,p_{\beta},
\end{equation}
where $g^{\alpha\beta}$ are the contravariant components of the metric tensor, and $p_\alpha$ is the canonical momentum conjugate to the coordinates $x^\alpha$. Both the orbit of the satellite, as well as its tick signals, share the same Hamiltonian, 
whose value is conserved along all trajectories. For the case of signal propagation it is null, and for the satellite, under 
our choice of signature, negative. We compute the trajectories by integrating the eight Hamilton equations for the 
generalized coordinates $x^\alpha$ and the momentum $p^{\alpha}$ with respect to proper time. Then at equal intervals of time 
in the spacecraft clock's frame, we find the signal path which traverses the spacetime to terminate at the receiver, which we 
place at an arbitrary location on the Earth's surface. Because space is curved, these paths are not globally straight, 
necessitating the solving of a boundary value problem in order to find the particular null trajectories which terminate 
at the Earth observer. Once we have the orbit solution, as well as the tick-propagation solutions along the orbit, we have 
the tick coordinate times of arrival. By taking the derivative with respect to the proper time of emission 
[Eq. \ref{crux}], we get the redshift. Further details on the calculation are described in the Appendix. The same method has been used successfully to calculate analogous effects in \cite{gc1, gc2} for stars in orbit around the Galactic-center supermassive black hole. 

In weak gravitational fields like the ones we encounter in the Solar System, the equations of general relativity reduce to those of Newtonian mechanics at order $\mathcal{O}(r^{-1})$ plus some small corrections at higher orders. Here $r$ is the distance that the satellite has from the origin, and is inversely proportional to the strength of the field. The relativistic perturbations are summarized in Table \ref{tab:Terms}, and described here.

\begin{itemize}
\item At $\mathcal{O}(r^{-3/2})$, there are two effects, both due to space curvature: namely its influence on the orbit and the light paths. The orbital effect is the same one that causes the orbit of Mercury to precess, and may be observable in extrasolar planets as well \cite{exoplanets_and_gr}.
But rather than simply measuring a cumulative precession over many orbits, a satellite that carries a clock would be sensitive to space curvature influencing the spacecraft trajectory over the course of a single orbit, especially near perigee.  This space curvature also bends the signal propagation paths. This light-path effect is well known as the Shapiro delay.   Both of these effects are measurable in binary pulsars \cite{pulsar2}. 
\item  At $\mathcal{O}(r^{-2})$, the frame-dragging of spacetime due to the Earth's spin enters the picture.  Frame-dragging has been measured around the Earth by Gravity Probe B via Lense-Thirring precession \cite{gravityprobeb}, and is expected to be detected shortly by LaReS \cite{lares} via frame-dragging-induced orbital precessional through precise orbit determination.
\item At $\mathcal{O}(r^{-5/2})$, spin-squared effects arise.  These are several effects that are proportional to the square of the Earth's spin. They have counterintuitive effects on both orbits and light paths, and though expected, have never been observed.
\end{itemize} 

\begin{figure*}
\begin{center}
\includegraphics[scale = 0.77, trim=0.2cm 0.2cm 0.2cm 0.3cm,clip=false]{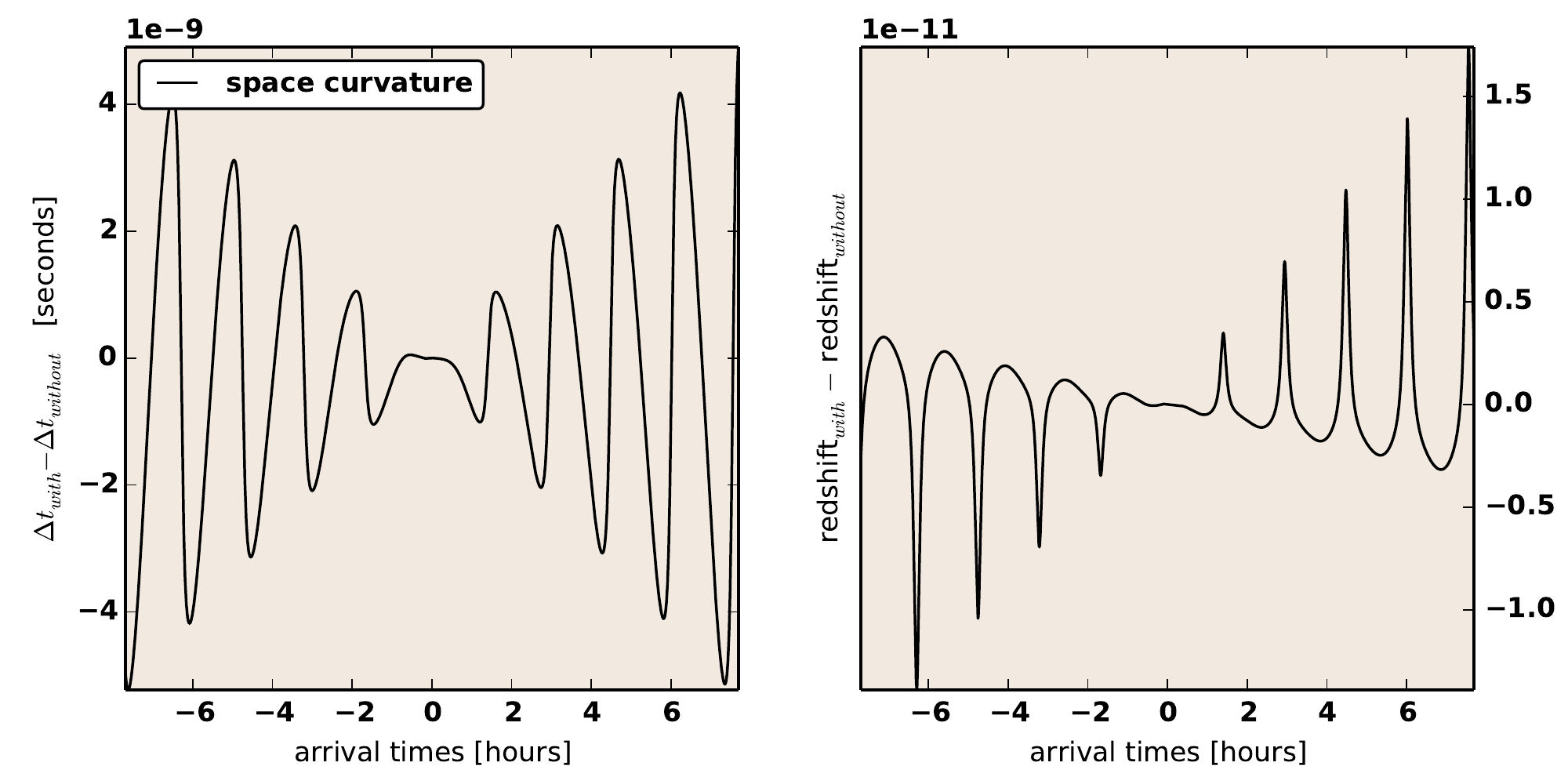}
\caption{\label{fig:ISS}
The orbital space-curvature (Schwarzschild) signals from a clock on a circular low Earth orbit. This effect is often parametrized through the PPN parameters $\beta$ and $\gamma$. The integration begins at $t=0$, at which time the satellite is at perigee, and runs forwards and backwards for five orbits in each direction. This is done with the effect turned off, and then turned on. The difference yields the signals shown here. At $t=0$, the two line up, so no relativity is seen. Because the metric terms (third block down on the second column in Table \ref{tab:Terms}) here affect the orbit, the effect builds up over many orbits. However, transient features also play a part. Effects that enter at a higher order than this one are significantly weaker, and are unlikely to be detectable on a circular orbit.}
\end{center}
\end{figure*}

\begin{figure}
\begin{center}
\includegraphics[scale = 0.6, trim=0.2cm 0.2cm 0.2cm 0.3cm,clip=true]{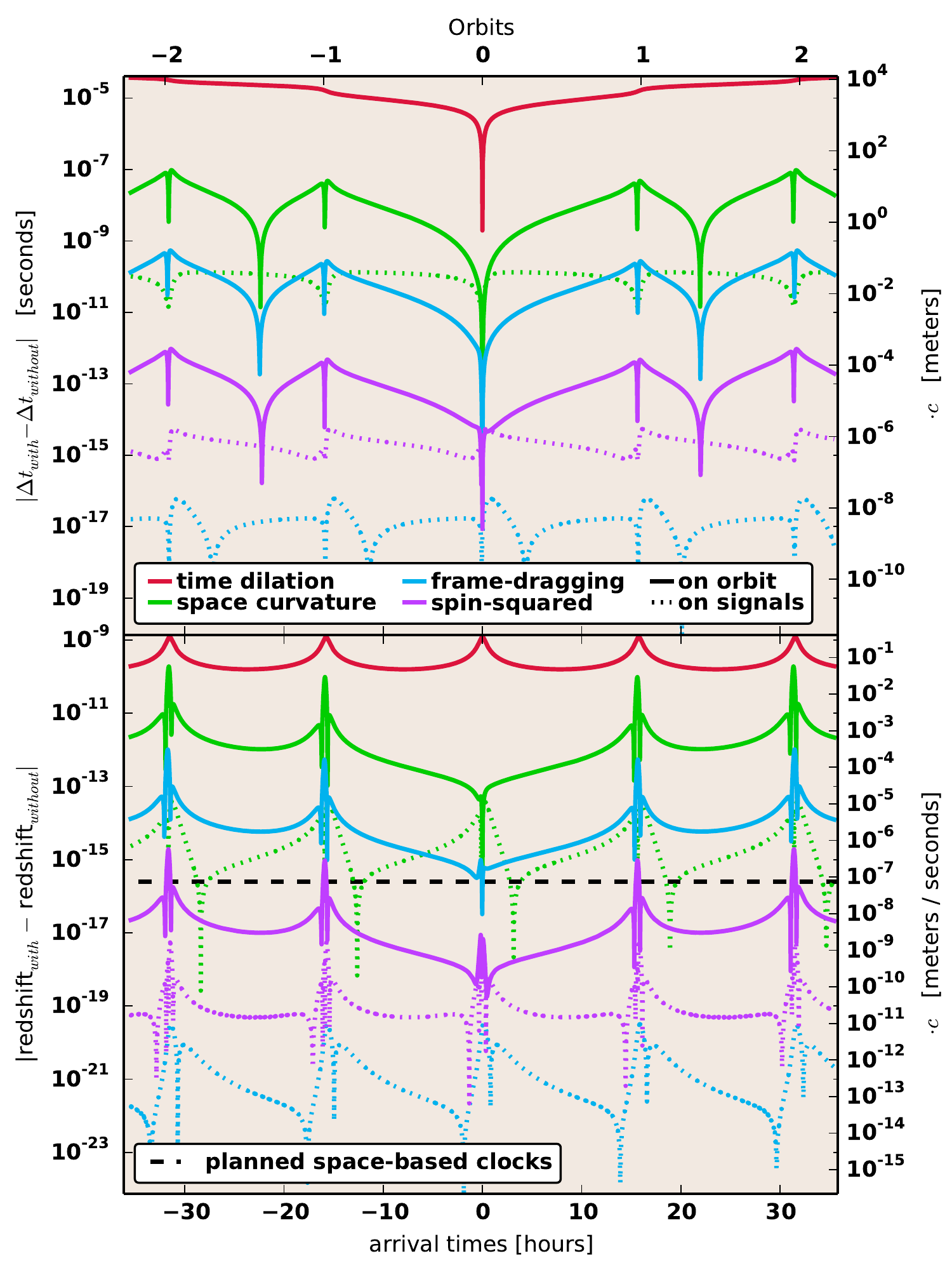}
\caption{\label{fig:all_signals}Relativistic timing signals (above) and associated redshift signals (below) of the setup described, for four and a half orbital periods. Solid curves correspond to orbit effects, while the dotted ones to signal propagation. The black dashed line corresponds to the $2\times 10^{-7}$ clock accuracy of planned\cite{stequest} space satellite missions. The right axes multiply the left ones by $c$. For the timing delay, this corresponds to the induced position shifts, and for the redshift perturbations, to the line-of-sight velocity perturbation. However, this interpretation is valid only for relativistic orbit effects - not those from tick signal propagation.}
\end{center}
\end{figure}

The geometry is described by the weak-field Kerr metric, a solution to the Einstein field equations, and relevant to the Earth's external field. Table~\ref{tab:Terms} shows the Hamiltonian expanded in powers of $1/r$ with the terms grouped according to the order in which they affect the dynamics. While the Hamiltonian for the satellite is identical to that for the photon signals, the order at which various terms enter the dynamics differs, due to the different behavior of the momenta. Any terms may be parametrized. For example, the $\mathcal{O}(r^{-3/2})$ space-curvature terms are popularly adjusted through $\beta$ and $\gamma$, part of the parametrized post-Newtonian (PPN) framework \cite{willEPPandPPN}, by introducing them as coefficients to the responsible Hamiltonian terms, although in this paper they are implicitly set to the Einstein values: unity. 

To calculate the signal on the redshift due to a particular relativistic effect, we first calculate the redshift with the corresponding Hamiltonian terms present; and then again, without. The difference yields the signal. We examine two orbits here. The first is a circular low Earth orbit, and the second eccentric.
\begin{enumerate}
\item \textbf{Circular orbit.} 
This circular orbit has semi-major axis $a = 6,800\, \rm{km}$, and an inclination of $51^\circ$ from the observer. For a circular orbit, the gravitational time dilation redshift is constant, implying a linear time-delay drift due to this effect. The next highest order relativistic effect is that due to space curvature. The time-delay and redshift signals due to space curvature on this orbit are plotted in Fig. \ref{fig:ISS} for 10 orbit periods. Higher-order effects are unlikely to be detectable on such an orbit because atmospheric drag severely restricts integration times. 

\item \textbf{Eccentric orbit.}
This orbit has semi-major axis is $a \approx 32,000\,\rm{km}$, and is eccentric with $e=0.77$. This satellite has an inclination relative 
to the Earth's spin plane of $s_\theta \approx 63^\circ,$ and longitude of the ascending node $s_\phi = 0^\circ$ (the full system has rotational symmetry about this direction). Figure \ref{fig:orientation} shows the setup. We choose to place the ground station on the Earth's equator - therefore with an inclination $I = -s_\theta$ with $\omega = 45^\circ$. (The observer's angular elements are defined relative to the orbital plane.)  The Earth's spin parameter is $s \equiv J_\oplus /M_\oplus  \approx 888$. The results for four and a half orbits of this geometrical configuration are shown in Fig. \ref{fig:all_signals}, which plots the magnitudes of the time-delay and corresponding redshift signals.  
\end{enumerate}

Both orbits are similar to future spacecraft clock missions. The circular orbit is based on the ACES mission, which will put an atomic clock on the 
International Space Station that communicates with the best Earth clocks available \citep{aces}. The eccentric case belongs to the proposed 
Space-Time Explorer and Quantum Equivalence (STE-QUEST) \cite{STE-QUEST-cqg, missionrequirementsdocument}, a medium class mission proposed by the European Space Agency. Relativistic effects on the latter are more pronounced by factors $\sim 10-100$. 
The peaks in the redshift due to relativistic orbit effects occur during perigee passage, where the satellite is moving fastest. The high 
perigee velocities offered by eccentric orbits significantly boost relativistic effects, as expected from the terms in the first column of 
Table~\ref{tab:Terms}. 

In both cases presented here, the receiving clock is assumed fixed and does not follow the Earth's rotation. Were we to include the receiver's motion, none of the relativistic signals would change, as its effect on the redshift is nonrelativistic, and so would be subtracted away. Its inclusion will be necessary for solving the inverse problem. Another simplification is that the Earth is penetrable by the tick-propagation signals, whereas in reality, portions of the integration period would miss data during occlusion. For clarity, we do not exclude these portions from the plots, however a single ground station would have $<50\%$satellite visibility, depending on the observer position and the orbit geometry. Missing data during occlusion do not affect the capability for performing long integration periods - important for letting cumulative relativistic effects build up.  Missions like STE-QUEST and ACES plan to have multiple ground stations, which will provide more complete coverage, although a single ground station would suffice. 

\section{Further redshift-influencing factors}
Multiple other nonrelativistic sources can be expected to influence the timing, which are not addressed here, yet will become important for the inverse problem.
\subsection{Variation of the fine-structure constant}

Since interactions between electron fields and photon fields are remarkably well understood through quantum electrodynamics (QED), an atomic clock in space offers more than timekeeping: it can test the equivalence principle for QED to remarkable accuracy. For example, a modulation of the clock's intrinsic frequency $\nu_0$ with the strength of the gravitational field $1/r$ would constitute a violation. This would imply that QED, a nongravitational theory, does not completely reduce to its conventional special relativistic limit in a local, freely falling frame. One way to approach a violation is by allowing the fine-structure constant $\alpha$ (or the charge on the electron) to depend on gravitational field strength.  Satellite timing experiments using clocks on eccentric orbits can test the gravitational field strength invariance of QED by simply promoting $\nu_0$ from a known quantity to a to-be-determined function. A natural way to examine QED-violating behavior is by letting $\nu_0\left(r\right) = \nu_\infty \left(1+\Xi\left(r\right) \right),$ and parametrize it with a constant $\xi$ via $\Xi\left( r\right) = \xi/r$. Assuming the intrinsic frequency can be recovered to the clock stability of $\sim10^{-16}$, then its values at apogee and perigee can constrain this possible violation of QED's local position invariance to $\Delta \xi/\xi \sim 10^{-7}$. Mechanisms which give rise to equivalence principle-violating behaviour in atomic clocks (and therefore affect redshift) are explored in \cite{EP_fail}.

\subsection{Solar System bodies}

 Gravitational influences beyond those discussed in this paper will affect the redshift. At accuracy levels of $10^{-16}$, significant Newtonian contributions from geophysical mass perturbations will become relevant.  For example, the $J_2$ value of the Earth  gives a correction of $10^{-3}$ relative to the Newtonian potential. Measurements of the Earth gravitational field have been performed \citep{EGM08} and can be used to remove these effects.
 
There are several other gravitational effects, due to the tidal fields of the Sun and Moon, that are similar in 
magnitude to the effects that are studied in this paper. For example, the Moon will also contribute a relativistic bending delay to the tick trajectories at the level of $\Delta z \sim 10^{-16}$. Solar System ephemerides allow extremely accurate modeling of 
the position of the Sun and the Moon, and the planets, which can be used to model tidal effects.

\subsection{Nongravitational forces}

A critical factor influencing the extent to which relativistic orbit effects may be recovered is the length of time over which the satellite is freely falling, i.e., geodesic. Nonballistic accelerations such as satellite reorientation routines, thrusting maneuvers, and atmospheric drag will limit integration times, suppressing the benefits offered from effects which accumulate.  A discussion on how atmospheric drag affects satellite orbits is given in \cite{flybyanomaly}, which suggests that a perigee altitude of $\sim 700$ km corresponds to drag accelerations as large as $10^{-6}\,\rm{m\,s}^{-2}$, although it drops exponentially with altitude. Solar radiation pressure generates accelerations of $\sim 10^{-7}\,\rm{m\,s}^{-2}$ assuming a spacecraft of mass of $2000\,\rm{kg}$ and an effective area of $10\,\rm{m}^2$. An accelerometer aboard the satellite would be insensitive to gravitational effects, but sensitive to nongravitational accelerations. Its measurements can then be used to correct for nongravitational accelerations. Measuring these accelerations to the level of the clock's accuracy would require an accelerometer with an accuracy of $\sim 10^{-12}\,\rm{m\,s}^{-2}$. The accelerometers used in the GRACE mission have an accuracy $\sim10^{-10}\,\rm{m\,s}^{-2}$ and an accuracy $\sim10^{-14}\,\rm{m\,s}^{-2}$ should be achieved by the LISA Pathfinder.  Signal accumulation may be necessary to resolve the frame-dragging orbit effect, and in particular to resolve the spin-squared effects. Signal propagation effects however are purely transient, and so their recovery can be expected to be only marginally affected by such lapses in falling freedom. Furthermore, these contributions affect the orbit with different periodicities and characteristic shapes, which should allow them to be separated from the desired terms when inverting the measurements of the timing residuals.

\section{Outlook and Challenges}

The approach taken in this paper generalizes trivially to similar timing experiments. For example, such a mission carried out around the Sun would benefit from field strengths a few orders of magnitude stronger. Furthermore, resolving the frame-dragging or the spin-squared signals would provide an independent measurement of the Sun's total angular momentum, heretofore measured reliably only from helioseismology \cite{iorio}. Another possibility would be to replace the ground-space clock pair with a single Earth-based clock, and let the satellite (whether orbiting the Earth or Sun) act as a mirror: either passively reflecting or actively retransmitting incoming ticks back to the terrestrial station, to be compared with the emitting clock. The physics relevant for such a mission is almost identical to that discussed in this paper.

The observational strategy will play a crucial role in the extent to which the relativistic perturbations discussed in this paper may be resolved.  The existing formalism for orbit determination \cite{2009A&A...504..653D} using the Global Navigation Satellite System for positioning in effect includes only the first four lines from Table~\ref{tab:Terms}.  Hence the orbit will need to be refined using the timing signals themselves.  Once measurements are taken, simultaneous fits to timing data via models which include a variety of both relativistic and nonrelativistic influences will provide precise orbit determination, and by doing so reveal information on Earth's exterior gravitational field at unprecedented accuracy levels.

Up to now, timing experiments in Earth's gravitational field have been focused on testing gravitational time dilation - a well-tested and understood consequence of the Einstein equivalence principle. Experiments of the type discussed in this paper probe higher-order terms of the gravitational field equations offering the exciting prospect of
testing a wide range of alternative theories of gravity. One class of such alternative theories are scalar-tensor theories, where the gravity action contains a scalar field in addition to the metric tensor of general relativity.
These theories are usually metric, and thus they respect the weak equivalence principle, yet they predict a $\gamma$ different from unity.  If the coupling function, unlike in Brans-Dicke theory, is not constant, the $\beta$-parameter also varies from unity. Therefore by testing the precession of the orbit and the Shapiro delay this class of theories can be tested. Similarly, the PPN parameters $\alpha_1$ and $\alpha_2$ can enter in the frame-dragging terms. In vector-tensor theories, for example, these are expected to deviate from zero - their general relativity value.

Up until now, the relativistic behavior of freely falling bodies was probed by various usually unrelated experiments, all in independent astrophysical systems. A different effect (or box in Table \ref{tab:Terms}) has always asked for a different experiment. Precise timing experiments of the type ``possible with next-generation space-based clock technology'' will be simultaneously sensitive to all these effects, as well as as-yet undetected ones, through the course of a single experiment. However, definitive statements regarding detectability can only be made by solving the mission-specific comprehensive inverse problem through realistic mock data generation, and Monte Carlo recovery of the parameters. 

\begin{acknowledgments}
R.A. thanks L. Tancredi for helpful discussion, and acknowledges support from the Swiss National Science Foundation. R.B. is grateful to M. Bondarescu for useful discussion, and acknowledges support from the Dr. Tomalla Foundation. We thank the referees for valuable, constructive reports.
\end{acknowledgments}

\def\aap{A\&A}

\bibliography{ms}

\def\disp{\displaystyle \vrule width 0pt height 24pt depth 20pt}
\def\desc#1#2{\vbox{\hsize=0.25\hsize
              \parindent=0pt \leftskip=0pt plus1fil \rightskip=\leftskip
              \centerline{\phantom{\bigg|}#1} \medskip
              \centerline{$\Delta t/t \sim r^{\,#2}$}}}

\begin{table*}[]\begin{center}
\begin{tabular}{|l l||c|c|} \hline 
 & & Satellite orbit & Signal propagation paths  \\
 \hline
&$\disp -\frac{p_t^2}{2}$  & static &\multirow{2}[4]{*}{Minkowski (straight)} \\
\cline{3-1}
& +$\disp\frac{p_{\phi}^2}{2r^2\sin^2\theta} + \frac{p_r^2}{2} + \frac{p_\theta^2}{2r^2}$ & \multirow{2}[6]{*}{\desc{Kepler (ellipse)}{-1/2}}&   \\
\cline{4-1}
& $\disp-\frac{p_t^2}{r}$ & & \multirow{2}[4]{*}{\desc{Shapiro delay}{-3/2}} \\
\cline{3-1} 
& $\disp-\frac{p_r^2}{r}$ &\multirow{2}[6]{*}{\desc{Schwarzschild}{-3/2}} &  \\
\cline{4-1}
&$\disp-\frac{2p_t^2}{r^2}$ & & \multirow{3}[10]{*}{\desc{Spin (both), Shapiro}{-5/2}} \\
\cline{3-1}
&$\disp-\frac{2s p_t p_\phi}{r^3}$ & \desc{Spin}{-2} & \\
\cline{3-1}
&$\disp+s^2 \left(\frac{p_r^2\sin^2\theta}{2r^2}-\frac{p_\theta^2\cos^2\theta}{2r^4} - \frac{p_\phi^2}{2r^4\sin^2\theta} \right.$ &\multirow{2}[8]{*}{\desc{Spin$^2$, Schwarzschild}{-5/2}} &  \\
\cline{4-1}
&$\disp\left. + \frac{p_t^2\cos^2\theta}{r^3}\right) -\frac{4p_t^2}{r^3}$ & & not included \\
\hline
\noalign{\medskip}
\end{tabular}
\end{center}
\caption{\label{tab:Terms} Hamiltonian terms for orbits and light paths, along with the
scaling of the fractional time-delay $\Delta t /t$ (redshift) with orbit size $r$. Refer to \cite{gc1} for the derivation. Note that we are using geometrized units $GM=c=1$.  To put $\Delta t$ and $r$ in different units, simply multiply by the appropriate powers of $GM$ and $c$ to fix the dimensions.}

\end{table*}

\newpage

\appendix

\section{Calculating the trajectories}\label{appendix}
While the form of the Hamiltonian found in Table \ref{tab:Terms} helps to provide an intuitive description of how relativity steers freely falling bodies, in practice the spherical coordinate system is less numerically stable than a pseudo cartesian one. This because the integration of photon paths which are almost straight is trivial in the latter. It is therefore convenient to canonically transform the Hamiltonian through
\begin{equation}
\begin{aligned}
x_\mu &= \left(t, r, \theta, \phi \right)
      \longrightarrow \left(t, \bf{x} \right) \\  
p_\mu &= \left(p_t, p_r, p_\theta, p_\phi \right)
      \longrightarrow \left(p_t, \bf{p} \right).  
\end{aligned}
\end{equation}
Under the generating function  
\begin{equation}
S = r\sin\theta\cos\phi \, p_x + r\sin\phi\sin\theta \, p_y
  + r\cos\theta \, p_z,
\end{equation}
the canonical momenta in the two bases are related by 
\begin{equation}
\begin{aligned}
p_r      & \equiv \frac{\partial S}{\partial r} =
\frac{{\bf x}\cdot {\bf p}^2}{r}, \\
p_\phi   & \equiv \frac{\partial S}{\partial \theta} =
\left(\bf{x} \times \bf{p}\right)_z, \\
p_\theta & \equiv \frac{\partial S}{\partial \phi} =
\frac{-1}{\sqrt{1-(z/r)^2}}
\left( p_z r - \frac zr \bf{x} \cdot \bf{p} \right).
\end{aligned}
\end{equation}
Inserting these into $H$, we have the Hamiltonian in
pseudo-Minkowskian coordinates.
\begin{eqnarray*} \label{manylines}
H &=& -\frac{p_t^2}{2} + \frac{\textbf{p}^2}{2} - \frac{p_t^2}{r} -\frac{({\bf x}\cdot{\bf p})^2}{r^3} - \frac{2 p_t^2}{r^2} - \frac{2 sp_t \, (xp_y-y_px)}{r^3}\\
  && -4\frac{p_t^2}{r^3} + \frac{s^2p_t^2}{r^3}\frac{x^2+y^2}{r^2}
 + \frac{s^2}{2}\frac{1}{r^6}\left(x^2+y^2 \right)
   ({\bf x}\cdot{\bf p})^2 \\
  && -\frac{s^2}{2}\frac{1}{r^4}\frac{1}{1-z^2/r^2}\left(p_zr -
    \frac zr \bf{x}\cdot \bf{p} \right)^2
   \left(1-\frac{x^2+y^2}{r^2} \right) \\ 
 &&  -\frac{s^2}{2}\frac{1}{r^2}\frac{\left(xp_y-yp_x\right)^2}{x^2+y^2}
\end{eqnarray*}
The Hamilton equations corresponding to this Hamiltonian are used to integrate both the satellite and the tick signal trajectories. However, because photon paths have a null inner product of the momentum, and satellite paths do not, different terms in \ref{manylines} enter the dynamics of each at different orders, hence the different groupings of terms in Table \ref{tab:Terms}. With the appropriate initial conditions, the satellite orbit can be calculated. To find the tick signal trajectories which originate on the orbit in equal intervals of proper time, and terminate at a specific Earthbound position (the observer), a boundary value problem must be solved. Were the tick signal trajectories straight, they could simply leave the satellite aimed in the direction of the observer. However, because the tick signal path momenta is not constant (they curve according to the terms in the rightmost column of Table \ref{tab:Terms}), the correct initial momentum must be calculated by shooting multiple times: optimising the initial momenta using the distance of the trajectories' termination point from the observer. This procedure is thoroughly detailed in \cite{gc1}.

The program which makes the clock orbit the Earth, and transmit tick signals to the Earth observer is available as supplementary material in the form of compilable source code. The kernel is written in the C language and relies on a few libraries\citep{gsl, nlopt}, while the interactive user front-end is written with Python's matplotlib\citep{matplotlib}. The interface provides a schematic of the geometry, and allows the user to adjust the orbital parameters, and choose which relativistic effects to include.  
\end{document}